\documentclass[prb,twocolumn,showpacs,superscriptaddress]{revtex4}
\usepackage{amssymb}
\usepackage{graphicx}
\usepackage{bm}

\input{tcilatex}

\begin{document}

\title{Conductance of the elliptically shaped quantum wire}
\author{S. N. Shevchenko and Yu. A. Kolesnichenko}
\affiliation{B.I.Verkin Institute for Low Temperature Physics and Engineering, 47 Lenin
Ave., 61103 Kharkov, Ukraine.}

\begin{abstract}
The conductance of ballistic elliptically shaped quantum wire is
investigated theoretically. It is shown that the effect of the curvature
results in strong oscillating dependence of the conductance on the applied
bias.
\end{abstract}

\pacs{73.20.Dx, 73.50.-h.}
\maketitle

\section{Introduction}

Recent advances in semiconductor physics and technology enabled the
fabrication and investigation of nanostructure devices which have some
important properties, such as small size, reduced dimensionality, relatively
small density of charge carriers and, hence, large mean free path (which
means that particles exist in the ballistic regime, so that scattering
processes can be neglected) and large Fermi wavelength $\lambda _{F}$. One
of mesoscopic systems of particular interest is quantum wire in which
particles are constrained to move along a one-dimensional curve due to
quantization of the transverse modes \cite{foot1}. And one of the numerous
important problems about that is the influence of the process of reducing
the dimensionality upon properties of the system.

Jensen and Koppe \cite{Jensen} and da Costa \cite{daCosta} has emphasized
that a low dimensional system, in general, has some knowledge of its
surrounding three-dimensional Cartesian space: the effective potential
arises from the mesoscopic confinement process which constrains particles to
move in domain of reduced dimensionality. Namely, it was shown that a
particle moving in a one- or two-dimensional domain is affected by
attractive effective potential \cite{daCosta}; for the first time this
result was obtained in Ref. \cite{Marcus} and later in Ref. \cite{Switkes}.
This idea was widely studied by several other authors (see Refs. \cite{Exner}%
$^{-}$\cite{Yu.A.} and, for example, Ref. \cite{Prinz} about experimental
realization of such systems).

It was also shown in Ref. \cite{Kugler} that the torsion of the twisted
waveguide affects the wave propagation in the waveguide independently of the
nature of the wave. In particular, the torsion of the waveguide results in
the rotation of polarization of light in a twisted optical fiber \cite{Chiao}%
. In Ref. \cite{Goldstone} the authors prove that in a waveguide, be it
quantum or electromagnetic one, exist bound states. And there are also some
references concerning the relation of the quantum waveguide theory to the
classical theory of acoustic and electromagnetic waveguides in Ref. \cite%
{Exner2}.

The effect of the curvature on quantum properties of electrons on a
two-dimensional surface, in a quantum waveguide, or in a quantum wire can be
observed by investigating kinetic and thermodynamic characteristics of
quantum systems \cite{Clark2}$^{-}$\cite{Yu.A.}. In this paper we propose to
use for this purpose measurements of the conductance $G$ of a quantum wire
and we show that the reflection of electrons from regions with variable
curvature results in non-monotonous dependence of the conductance on the
applied bias.

In Ref. \cite{Switkes} the Schr\"{o}dinger equation on the elliptically
shaped ring was solved numerically in order to get the eigenvalue spectrum
of a particle confined to the ring. The authors demonstrated that the
behavior of a quantum mechanical system confined by the rectangular well
potential to a narrow ring in the limit when its width $\gamma $\ tends to
zero is analogous to the straight line motion with effective potential 
\begin{equation}
V_{eff}=-\frac{\hbar ^{2}}{8mR^{2}},  \label{Veff}
\end{equation}%
where $R$ is the radius of curvature. Later, in Ref. \cite{Magarill} the
electron energy spectrum in an elliptical quantum ring was considered in
connection with the persistent current; the authors have concluded that the
effective potentials $V_{eff}$ are different for different confining
potentials even in the limit when $\gamma $\ tends to zero. This conclusion
is in contradiction with results of some other papers \cite{daCosta}, \cite%
{Exner2}. So, in this paper we elaborate the problem; we investigate the way
of derivation of the one-dimensional Schr\"{o}dinger equation in order to
understand deeper how constraining potential affects particle motion along
the curve $C$; we demonstrate the consistency with previous results \cite%
{daCosta}: the effective potential is universal for different confining
potentials and depends only on the curvature (see Eq.(\ref{Veff})).

\begin{figure}[t]
\includegraphics[width=6.0cm]{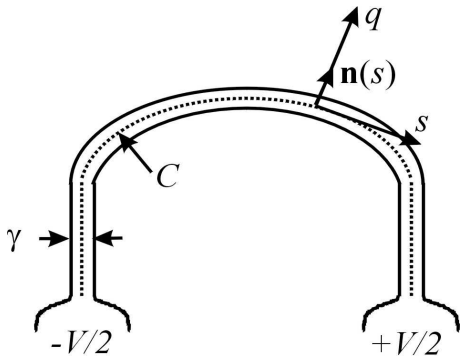}
\caption{Elliptically shaped quantum wire.}
\label{Scheme}
\end{figure}

In Sec.2 we study derivation of the one-dimensional Schr\"{o}dinger equation
starting from the two-dimensional Schr\"{o}dinger equation which describes a
non-relativistic electron moving in a plane \cite{foot2} and being the
subject to the confining potential $V_{\gamma }$. Further, in Sec.3, we
apply this results to consider theoretically the conductance of the quantum
wire which consists of two linear parts and one elliptically shaped part
between them; the wire is connected to two conducting reservoirs at
different voltages (see Fig.\ref{Scheme}). And in Sec.4 we discuss the
influence of the curvature on the conductance. 

\section{Schr\"{o}dinger Equation}

In this section we will follow the approach proposed in Ref. \cite{daCosta}.
Let us consider an electron with effective mass $m$ moving in a quantum wire
along a curve $C$ which is constructed by a prior confinement potential $%
V_{\gamma }$. For the sake of simplicity, we start from two-dimensional
motion. We introduce the orthonormal coordinate system \cite{foot3} $(s,q)$,
where $s$ is the arc length parameter and $q$ is the coordinate along the
normal $\overrightarrow{n}=\overrightarrow{n}(s)$ to the reference curve $C$%
. Then the curve $C$ is described by a vector valued function $%
\overrightarrow{r}(s)$ of the arc length $s$. So, the position in proximity
to the curve $C$ is described by

\begin{equation}
\overrightarrow{R}(s,q)=\overrightarrow{r}(s)+q\cdot \overrightarrow{n}(s).
\end{equation}

To obtain a meaningful result the particle wave function should be
''uniformly compressed'' into a curve, avoiding in this way the tangential
forces \cite{daCosta}, \cite{Switkes}, \cite{Magarill}. So, we consider $%
V_{\gamma }$ to be dependent only on the $q$ coordinate which describes the
displacement from the reference curve $C$, this means that points with the
same $q$ coordinate but different $s$ coordinates (which describe the
position on curve $C$) have the same potential. This potential contains
small parameter $\gamma $\ so that the potential increases sharply for every
small displacement in the normal direction; $\gamma $ is a characteristic
width of the potential well $V_{\gamma }$ (the simplest examples of such a
potential are the rectangular well potential and the parabolic-trough
potential, albeit the real potential would likely be the combination of
both). So, the small parameter of the problem is $\gamma /R\ll 1$ \cite%
{Exner}.

The motion of the electron obeys the time-independent Schr\"{o}dinger
equation which has the form

\begin{equation}
-\frac{\hbar ^{2}}{2m}\triangle _{s,q}\psi +V_{\gamma }(q)\psi =\varepsilon
\psi ,  \label{Init_Schr_Eq}
\end{equation}
\newline
where the Laplacian is

\begin{equation}
\triangle _{s,q}=\frac{1}{h}\frac{\partial }{\partial s}\frac{1}{h}\frac{%
\partial }{\partial s}+\frac{1}{h}\frac{\partial }{\partial q}h\frac{%
\partial }{\partial q},  \label{Laplacian}
\end{equation}
with $h$ being the Lam\'{e} coefficient (corresponding to the longitudinal
coordinate $s$) which depends thanks to Frenet equation upon the curvature $%
k=k(s)$:

\begin{equation}
h=1-k(s)\cdot q.
\end{equation}

In order to eliminate the first-order derivative with respect to $q$ from
Eq.(\ref{Init_Schr_Eq}) we naturally introduce \cite{foot4} new wave
function $\widetilde{\psi }:$

\begin{equation}
\widetilde{\psi }(s,q)=\sqrt{h}\psi (s,q),  \label{psi_waved}
\end{equation}%
which is, as a matter of fact, the wave function introduced in Ref. \cite%
{daCosta} which is normalized so that 
\begin{equation}
\int dsdq\left\vert \widetilde{\psi }(s,q)\right\vert ^{2}=1.
\label{normalization}
\end{equation}%
And then the Schr\"{o}dinger equation Eq.(\ref{Init_Schr_Eq}) yields

\begin{equation}
-\frac{\hbar ^{2}}{2m}\left( \frac{\partial }{\partial s}\frac{1}{h^{2}}%
\frac{\partial }{\partial s}+\frac{\partial ^{2}}{\partial q^{2}}\right) 
\widetilde{\psi }+V_{eff}(s,q)\widetilde{\psi }+V_{\gamma }(q)\widetilde{%
\psi }=\varepsilon \widetilde{\psi },  \label{Eq in s and q}
\end{equation}
where

\begin{equation}
V_{eff}(s,q)=-\frac{\hbar ^{2}}{2m}\left( h^{-2}\frac{k^{2}}{4}+\frac{q}{2}%
h^{-2}\frac{d^{2}k}{ds^{2}}+\frac{5q^{2}}{4}h^{-4}\left( \frac{dk}{ds}%
\right) ^{2}\right) ,
\end{equation}%
which is in agreement with Refs. \cite{Exner}, \cite{Clark2}.

One should be careful with Eq.(\ref{Eq in s and q}) in order to avoid
mistakes found in literature \cite{Clark}, \cite{Magarill}. First, we can
not decompose this equation, which contains terms which are functions of
both $s$ and $q,$ into two equations introducing $\widetilde{\psi }%
(s,q)=\chi _{n}(q)\chi _{t}(s)$ as in Ref. \cite{Clark}, where authors have
obtained the Eq.(31) for $\chi _{t}(s)$ with coefficients depending on the $%
q $ variable. To understand another mistake \cite{Magarill}, consider Eq.(%
\ref{Eq in s and q}) within the perturbation theory on small parameter $%
\gamma $ [comparatively with $R$] (see also in Ref. \cite{Exner2}). Expand $%
h^{-2}\ $and\ $V_{eff}$ in series in $q\lesssim \gamma ,$ writing down
explicitly the zeroth term:

\[
h^{-2}=1+\sum_{l=1}^{\infty }f_{l}(s)q^{l}, 
\]
\[
V_{eff}(s,q)=-\frac{\hbar ^{2}}{2m}\left( \frac{k^{2}(s)}{4}%
+\sum_{l=1}^{\infty }y_{l}(s)q^{l}\right) . 
\]
Then Eq.(\ref{Eq in s and q}) can be rewritten:

\begin{equation}
\left( \widehat{H}_{0}+\widehat{V}\right) \widetilde{\psi }=\varepsilon 
\widetilde{\psi },  \label{Eq with pert}
\end{equation}
where

\begin{equation}
\widehat{H}_{0}=-\frac{\hbar ^{2}}{2m}\left( \frac{\partial ^{2}}{\partial
s^{2}}+\frac{\partial ^{2}}{\partial q^{2}}\right) -\frac{\hbar ^{2}}{2m}%
\frac{k^{2}(s)}{4}+V_{\gamma }(q),  \label{H0}
\end{equation}

\begin{equation}
\widehat{V}=\frac{\hbar ^{2}}{2m}\sum_{l=1}^{\infty }q^{l}\left( -\frac{%
\partial }{\partial s}f_{l}(s)\frac{\partial }{\partial s}+y_{l}(s)\right) ,
\label{V_pert}
\end{equation}
we notice that $\widehat{V}$ is a second order in the variable $s$
differential operator. The solution of Eq. (\ref{Eq with pert}) is

\[
\widetilde{\psi }=\widetilde{\psi }^{(0)}+\sum_{l=1}^{\infty }\widetilde{%
\psi }^{(l)}, 
\]
where $\widetilde{\psi }^{(l)}\sim \gamma ^{l}$ and $\widetilde{\psi }^{(0)}$
corresponds to the zeroth-order approximation problem: $\widehat{H}_{0}%
\widetilde{\psi }^{(0)}=\varepsilon \widetilde{\psi }^{(0)}$. This equation
can be decomposed by separating the wave function: $\widetilde{\psi }%
(s,q)=\eta (q)\chi (s)$: 
\begin{equation}
-\frac{\hbar ^{2}}{2m}\frac{d^{2}}{dq^{2}}\eta +V_{\gamma }(q)\eta =E_{t}\eta
\label{transvEq}
\end{equation}
and 
\begin{equation}
-\frac{\hbar ^{2}}{2m}\frac{d^{2}}{ds^{2}}\chi +V_{eff}(s)\chi =E_{l}\chi ,
\label{longitEq}
\end{equation}
where $V_{eff}(s)$ is given by Eq.(\ref{Veff}); $\varepsilon =E_{t}+E_{l}$; $%
R=k(s)^{-1}$ is the curvature radius (in the next section we omit the
subscript ''$l$'', identifying energy $E$ with its longitudinal component $%
E_{l}$). Eq.(\ref{transvEq}) describes the confinement of the electron to a $%
\gamma $-neighborhood of the curve $C$ and Eq.(\ref{longitEq}) describes the
motion along the $s$ coordinate (along the curve $C$). In fact, Eq.(\ref%
{longitEq}) is a conventional one-dimensional Schr\"{o}dinger equation for
an electron moving in the $s$-dependent potential $V_{eff}(s)$; the latter
connects the geometry and the dynamical equation. The origin of this
potential is in the wavelike properties of the particles; $V_{eff}$ is
essential for not large $R/\lambda _{F}.$ We underline that the effective
potential (Eq.(\ref{Veff})) in the zeroth-order approximation in $\gamma /R$
is not dependent on the method of \ ''one-dimensionalization'', i.e. on the
choice of $V_{\gamma }(q)$ (compare this conclusion with the one derived in
Ref. \cite{Magarill}).

We also note that if we have started from the three-dimensional equation of
motion we would obtain an additional effective potential which in the planar
situation is zero \cite{daCosta}.

\section{Conductance}

The conductance $G$ of quantum contacts can be related to the transmission
probability $T(E)$\ by Landauer's formula \cite{Landauer}. At zero
temperature and finite voltages $V$ it takes the form

\begin{equation}
G=\frac{G_{0}}{2}[T(E_{F}+\frac{eV}{2})+T(E_{F}-\frac{eV}{2})]  \label{G}
\end{equation}
where $G_{o}=2e^{2}/h,$ $E_{F}$ is the Fermi energy. Two terms in this
equation correspond to two electronic beams moving in opposite directions
and differing in bias energy. So, we are interested in the transmission
probability $T(E)$ with $E$ being an electron energy.

In this section we consider the curve $C$ to consist of three ideally
connected parts (see Fig.\ref{Scheme}): (i) linear ($s<0$), (ii) elliptical (%
$0<s<l$, $l$ is half of the ellipse's perimeter), and (iii) one more linear
domain ($s>l$). We consider wave functions in the regions (i) and (iii) to
be plane waves $\psi _{1}=e^{ik_{1}s}+re^{-ik_{1}s},$ $\psi
_{3}=te^{ik_{1}s} $, where $k_{1}=\sqrt{2mE/\hbar ^{2}}$ is the wave vector
and $t$ and $r$ are the transmission and reflection coefficients; the
transmission probability is given by $T=\left| t\right| ^{2}$. The wave
function $\psi _{2}\equiv \chi $, where $\chi $\ is the solution of Eq.(\ref%
{longitEq}), with the effective potential given by Eq.(\ref{Veff}). The \
curvature can be written most simply in the elliptical $v$ coordinate \cite%
{Whittaker} which is defined with its Lam\'{e} coefficient

\begin{equation}
H=ds/dv=a\sqrt{1-e^{2}\cos ^{2}v},  \label{H}
\end{equation}
where $e$ is the eccentricity of an ellipse and $a$ is the length of its
major semiaxis; we apply $v(s=0)=0$. Then the effective (geometrical)
potential (see Eq.(\ref{Veff})) can be written as

\begin{equation}
V_{eff}(s)=-\frac{\hbar ^{2}}{8ma^{2}}\frac{1-e^{2}}{(1-e^{2}\cos ^{2}v)^{3}}%
,  \label{V}
\end{equation}
which is in agreement with Ref.\cite{Switkes}.

We introduce new wave function

\begin{equation}
\xi (v(s))=\chi (s)/\sqrt{H},  \label{ksi}
\end{equation}
for which the equation takes the form (see Eqs. (\ref{longitEq}), (\ref{H})-(%
\ref{ksi})):

\begin{equation}
\frac{d^{2}}{dv^{2}}\xi +\left[ \frac{2ma^{2}}{\hbar ^{2}}E\cdot g(v)+U(v)%
\right] \xi =0,  \label{Hill's_eq}
\end{equation}
\begin{equation}
U(v)=\frac{5}{4}\frac{1-e^{2}}{g^{2}}-\frac{1-e^{2}/2}{g}-\frac{e^{4}}{16}%
\frac{\sin ^{2}2v}{g^{2}},  \label{U}
\end{equation}
where $g=H^{2}/a^{2}=1-e^{2}\cos ^{2}v.$ Eq.(\ref{Hill's_eq}) is the Hill's
equation with $\pi $-periodic coefficients; fundamental system of its
solutions is \cite{Kamke}

\begin{equation}
\xi _{\pm }=e^{\pm i\mu v}y(\pm v),  \label{sol of Hill's Eq}
\end{equation}
where $y(v)$ is a $\pi $-periodic function, $\mu $ is the characteristic
exponent. Then (see Eqs.(\ref{ksi}) and (\ref{sol of Hill's Eq})):

\begin{equation}
\chi =C_{1}e^{i\mu v}\widetilde{y}(v)+C_{2}e^{-i\mu v}\widetilde{y}(-v),
\label{hi}
\end{equation}
(here $\widetilde{y}(v)\equiv \sqrt{H}y(v)$).

Thus, we know what wave functions are like and now we are interested in $%
T=\left| t\right| ^{2}$ which describes transmission over potential well
(see Eq.(\ref{V})). Then we make use of the conditions of continuity of the
wave function and of its derivative, so we obtain the system of four
equations which is similar to one given in Ref. \cite{LL}; and so is the
result: 
\begin{equation}
T=\left[ 1+\frac{1}{4}\left( \kappa -\frac{1}{\kappa }\right) ^{2}\sin
^{2}\pi \mu \right] ^{-1}  \label{T}
\end{equation}
where we denoted 
\begin{equation}
\kappa =-\frac{i}{ak_{1}\sqrt{1-e^{2}}}\left( \frac{\xi _{+}^{\prime }}{\xi
_{+}}\right) _{v=0}.
\end{equation}
(To get Eq.(\ref{T}) we applied that $\mu $ and $\kappa $ are real, which is
straightforward to proof.)

\section{Results and discussion}

To understand how the conductivity $G$ depends on the bias $eV$ and the
geometry we need to find the solution of Hill's equation Eq.(\ref{Hill's_eq}%
). We did this numerically as well as within the perturbation theory for an
ellipse which is close to a circle (i.e. $e^{2}\ll 1$); we found that both
are in good agreement for $e<1/2$. In the zeroth-order approximation in $%
e^{2}$ (i.e. for $e=0$, the case of a circular arc) we have $\mu _{0}=ak_{2}$
and $\kappa _{0}=k_{2}/k_{1},$ where $k_{2}=\sqrt{2mE/\hbar ^{2}+1/4a^{2}}$
(see also in Ref. \cite{Yu.A.}). This means that one can observe
oscillations in $G(V)$ dependence provided $a\gtrsim \lambda _{F}$ and the
amplitude of these oscillations is enough small.

The first-order approximation of the perturbation theory (for $a>\lambda
_{F} $) yields:

\begin{equation}
\mu \simeq \mu _{0}+e^{2}\mu _{1},\kappa \simeq \kappa _{0}+e^{2}\kappa _{1};
\label{mu_and_kappa}
\end{equation}
with

\begin{equation}
\mu _{1}=-ak_{1}^{2}/4k_{2}\equiv -\mu _{0}/4\kappa _{0},  \label{mu_1}
\end{equation}

\begin{equation}
\kappa _{1}=k_{1}/[4k_{2}((ak_{2})^{2}-1)].  \label{kappa_1}
\end{equation}

\begin{figure}[t]
\includegraphics[width=7.0cm]{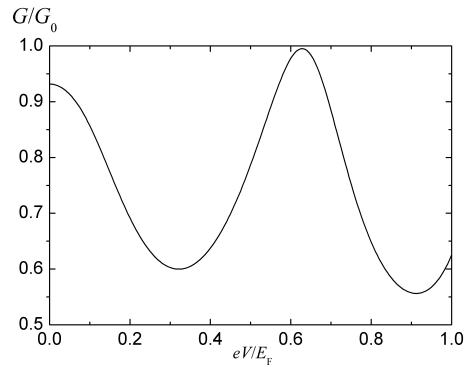}
\caption{Conductance as a function of the bias $G=G(eV)$ at $e=0.99$, $a=10%
\protect\lambda _{F}$ (at the same value of $a$ but $e=0$ the amplitude $%
\Delta G/G_{0}$ is of order of $10^{-5}$).}
\label{G(eV)}
\end{figure}

\begin{figure}[t]
\includegraphics[width=7.0cm]{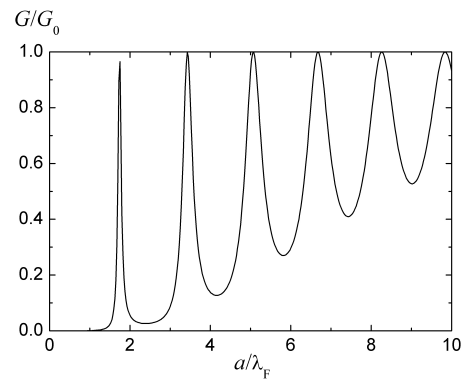}
\caption{Conductance as a function of the length of the major semiaxis\ $%
G=G(a)$ at $e=0.99$, $V=0$.}
\label{G(a)}
\end{figure}

Then we solve Hill's equation Eq.(\ref{Hill's_eq}) numerically. The
characteristic exponent $\mu $ is defined via the solution of Eq.(\ref%
{Hill's_eq}) with the initial conditions: $\xi _{1}(0)=1,$ $\xi _{1}^{\prime
}(0)=0$, and then $\mu $ is the solution of the equation$\ \xi _{1}(\pi
)=\cos \pi \mu $ (see in Ref. \cite{Kamke}). More challenging is to find $%
\xi _{+}$ (see Eq.(\ref{sol of Hill's Eq})), which can be formulated as the
boundary value problem with Eq.(\ref{Hill's_eq}) and boundary conditions $%
\xi _{2}(0)=0,$ $\xi _{2}(\pi )=\sin \pi \mu $ (where $\xi _{2}(v)=\func{Im}%
\xi _{+}(v))$. We introduce $\xi _{3}(v)=\xi _{2}(v)/\xi _{2}^{\prime }(0)$,
and so we have initial conditions problem for $\xi _{3}(v)$ ($\xi _{3}(0)=0,$
$\xi _{3}^{\prime }(0)=1$), the solution of which gives what we seek for to
define $\kappa $: $\left( \xi _{+}^{\prime }/\xi _{+}\right) _{v=0}=\xi
_{2}^{\prime }(0)=\sin \pi \mu /\xi _{3}(\pi )$. The results of the
described procedure are numerically plotted at Figs.\ref{G(eV)} and \ref%
{G(a)} which is for enough elongated ellipse with $e=0.99$ (so that $a/b=7,$
where $a$ and $b$ are the lengths of its major and minor semiaxes
respectively). We note about Fig. \ref{G(a)} that, being restricted by the
condition $R\gg \gamma $, we should not tend $a$ to $0$, namely, we may
suppose $R\gg \gamma $\ for $a/\lambda _{F}\sim 10$ but may not for $%
a/\lambda _{F}\lesssim 1$ [for $e$ close to $1$]. We also note that Eq.(\ref%
{G}) is, strictly speaking, correct for $eV$ small compared with $E_{F}$ and
describes $G(V)$ \ dependence for $eV\sim E_{F}$ qualitatively. We conclude
that $e$ close to $1$ increases significantly oscillations in comparison
with $e=0$ case; the amplitude of oscillations in $G=G(V)$ dependence is
defined by the value of $a/\lambda _{F}$. 

In summary, we have rederived the quantum-mechanical effective potential
induced by curvature of one-dimensional quantum wire. We have shown that for
any confining potential $V_{\gamma }$, depending only on the displacement $q$
from the reference curve $C$, this effective potential is universal: it does
not depend on the choice of $V_{\gamma }$ and is given by Eq.(\ref{Veff}).
And then we have studied the effect of the curvature in the zeroth-order
approximation in the width of the wire on the conductance of an ideal
elliptically shaped quantum wire. It has been shown, in particular, that,
due to the effect of the curvature, dependence of the conductance $G(V)$ on
the applied bias changes drastically. So, the effect of the curvature can be
observed by measuring the conductance of a quantum wire. On the other hand,
one can change the characteristics of the quantum wire, such as conductance,
setting its size, shape, or applied bias.

One of the authors (S.N.S.) would like to thank Prof. I.D. Vagner for his
warm hospitality during the stay at Grenoble High Magnetic Field Laboratory
(France) where the part of this work was done. We also thank Prof. A.M.
Kosevich for critical discussion of the manuscript.


\begin{thebibliography}{99}
\bibitem{foot1} We study here only one-channel wire when only the lowest
subband is occupied.
\bibitem{foot2} We consider here only flat curves and we refer the reader
who is interested in the effect of the torsion to Ref. \cite{Clark}.
\bibitem{foot3} The advantages of establishing from the very beginning the $%
(s,q)$ coordinate system are that it allows the most general consideration
and that [because of the diagonal structure of the metric tensor] we can in
the zeroth-order approximation in the width of the quantum wire decompose
the dynamical equation of motion into two.
\bibitem{foot4} We do this to eliminate terms of the form $f(q)\partial
/\partial q$ which in Ref. \cite{Jensen} were called \textquotedblright
dangerous terms\textquotedblright . We cannot apply $f(q)=f(0)$ since $%
f(q)\partial /\partial q\simeq \lbrack f(0)+q\cdot df(0)/dq]\partial
/\partial q$: although $q\sim \gamma $ but $\partial /\partial q\sim \gamma
^{-1}$ and, so, the second term in the brackets is $\sim \gamma ^{0}$ and
that is the order of terms we are interested in below.
\end{thebibliography}
\end{document}